\documentclass[aps,twocolumn,groupedaddress,superscriptaddress,showpacs,floats,amssymb]{revtex4}

\usepackage{amsmath}
\usepackage{amsfonts}
\usepackage{amssymb}
\usepackage{euscript}
\usepackage{bm}

\begin{document}

\title{Vortex-Phonon Interaction
in the Kosterlitz-Thouless Theory}

\author{Evgeny Kozik}
 \affiliation{Department of Physics, University of
Massachusetts, Amherst, MA 01003}
\author{Nikolay Prokof'ev}
\affiliation{Department of Physics, University of Massachusetts,
Amherst, MA 01003}
 \affiliation{Russian Research Center
``Kurchatov Institute'', 123182 Moscow, Russia}
\affiliation{Department of Physics, Cornell University, Ithaca, NY
14850}

\author{Boris Svistunov}
\affiliation{Department of Physics, University of Massachusetts,
Amherst, MA 01003} \affiliation{Russian Research Center
``Kurchatov Institute'', 123182 Moscow, Russia}

\begin{abstract}
The ``canonical" variables of the Kosterlitz-Thouless
theory---fields $\Phi_0({\bf r})$ and $\varphi({\bf r})$, generally
believed to stand for vortices and phonons (or their XY equivalents,
like spin waves, etc.) turn out to be neither vortices and phonons,
nor, strictly speaking, {\it canonical} variables. The latter fact
explains paradoxes of (i) absence of interaction between $\Phi_0$
and $\varphi$, and (ii) non-physical contribution of small vortex
pairs to long-range phase correlations. We resolve the paradoxes by
explicitly relating $\Phi_0$ and $\varphi$ to canonical vortex-pair
and phonon variables.
\end{abstract}


%
%
%
%
%
%
%
%
%
%
%

\maketitle

Three decades ago, Kosterlitz and Thouless  developed an accurate
renormalization-group description of what is now called a
Berezinskii-Kosterlitz-Thouless (BKT) transition
\cite{Berezinskii, KT}---a phase transitions in a wide class of
two-dimensional systems characterized by short-range interactions
and global U$(1)$ symmetry. In accordance with the Mermin-Wagner
theorem \cite{MerminWagner}, such systems can not exhibit
long-range order at any finite temperature. Instead, the
low-temperature phase features divergent long-wave fluctuations
resulting in a power-law decay of phase correlations at large
distances \cite{Berezinskii}. Kosterlitz and Thouless revealed the
importance of configurations with point-like topological defects
or topological charges, such as Coulomb charges in plasma,
dislocations in crystals, vortices in superfluid and spin systems,
etc. At low temperature, the defects exist only in the form a
dilute gas of bound pairs of opposite topological charges. At
higher temperature, pairs of large separation become more probable
and eventual pair dissociation at critical temperature destroys
the algebraic long-range order. Extensive theoretical work
\cite{Berezinskii, KT, Kosterlitz, Jose, Nelson_Kosterlitz,
Ambegaokar} provides a complete quantitative description of
critical properties in such systems. The theory is corroborated by
comprehensive experimental studies of $^4$He films \cite{Reppy,
Rudnick, Hallock81, Hallock82} and superconducting Josephson
arrays \cite{Resnick}. Recent advances in the area of ultra-cold
gases have made it possible to render BKT transition in optical
lattices \cite{Trombettoni_etal}.

The Kosterlitz-Thouless (KT) theory starts with a generic effective
action  \cite{KT}
\begin{equation}
A[\Phi]\; =\;  K_0 \int \bigl| \nabla \Phi \bigr|^2 d^2r\; .
\label{A}
\end{equation}
For definiteness, we consider the case of a superfluid film, in
which the field $\Phi(\mathbf{r})$ has the meaning of the velocity
potential, $\mathbf{v} = (1/m) \nabla \Phi$ (we set $\hbar = 1$),
and $K_0=n_0/2m T$, where $m$ is the mass of a particle and $n_0$
is the ``bare'' superfluid number density obtained by averaging
out microscopic fluctuations up to some mesoscopic scale $l_0$.
Hence, $n_0 \equiv n_0(l_0)$. The field $\Phi$ is then split,
\begin{equation}
\Phi =\Phi_0 + \varphi\; , \label{Phi}
\end{equation}
into a singular part $\Phi_0$, containing all the topological
defects, and a regular
part $\varphi$.

By definition, $\Phi_0$ satisfies the non-zero velocity circulation
condition
\begin{equation} \oint_{\mathcal{C}_j}
\nabla \Phi_0(\mathbf{r}) \cdot d \mathbf{r}\,  =\,  2 \pi l_j,
\label{circulation}
\end{equation}
where $\mathcal{C}_j$ is a contour enclosing only the $j$-th
defect and $l_j$ are integers, while $\nabla \varphi$ is
circulation-free. The next crucial step is to require \cite{KT}
\begin{equation}
\Delta \Phi_0(\mathbf{r})\, =\, 0\;  \label{Laplacian_Phi0}
\end{equation}
(except for the isolated points of defects). The standard
motivation of Eq.~(\ref{Laplacian_Phi0}) is that it guarantees
that $\Phi_0$  minimizes the action when $\varphi \equiv 0$. The
definitions of $\Phi_0$ and $\varphi$ thus become unambiguous and,
most importantly, the action takes the form of two independent
terms:
\begin{equation}
A[\Phi]\; =\;  A[\Phi_0]\, + \, A[\varphi] \; . \label{AA}
\end{equation}
At this point, one conjectures that $\Phi_0$ and $\varphi$
correspond to vortices and phonons (spin waves, etc.), respectively.
This identification, which might seem to be quite natural---or at
least merely terminological and mathematically irrelevant---is not
that innocent. If the two fields are not canonical vortices and
phonons, then one faces a problem of {\it justifying} writing the
partition function in the form \cite{KT}
\begin{equation}
Z\, \propto \,   \int \exp \left\{- A[\varphi]\right\} \mathcal{D}
\varphi \int \exp \left\{- A[\Phi_0] \right\} \prod_{j=1}^N d^2
r_j \; , \label{Z_kt}
\end{equation}
where ${\bf r}_j$ is the position of the $j$-th defect and $N$ is
the number of defects. This expression should also include the
Jacobian of the transformation from canonical variables.
Remarkable agreement between the Kosterlitz-Thouless theory and
experimental data suggests that the Jacobian is unimportant, but
without explicitly demonstrating this fact the theory is
incomplete.

Apart from Jacobian, there is also an issue of peculiar
``collective" behavior of formally independent fields $\Phi_0$ and
$\varphi$. Asymptotic long-range phase correlations in a superfluid
are due to phonons. The corresponding action in terms of the genuine
phonon field $\tilde{\varphi}$ is
\begin{equation}
A_{\tilde{\varphi}} = \frac{n_s}{2mT} \int \bigl| \nabla
\tilde{\varphi} \bigr|^2 d^2r \; , \label{A_varphi}
\end{equation}
with $n_s$ the macroscopic superfluid density. Would $\varphi$ stand
for phonons, we were to identify its long-wave harmonics with
$\tilde{\varphi}$. This, however, would imply $n_0(l_0) = n_s$,
which is definitely not the case since $\Phi_0 \neq 0$ at the length
scale $l_0$. This paradox can be formulated as an observation that
it is impossible to renormalize the sound velocity by vortex pairs
without the {\it phonon-vortex coupling}. The only logical solution
is then that $\varphi$ is {\it not} a phonon field.

Another paradoxical circumstance is associated with interpreting
$\Phi_0$ as a {\it purely} vortex field. In a 2D superfluid, all
vortices are bound in microscopic pairs and one would not expect
them to be {\it directly} observable in long-range correlation
properties. The only physical way for vortex pairs to manifest
themselves at the macroscopic scale is to renormalize the
superfluid density. However, the requirement
(\ref{Laplacian_Phi0}) implies that vortex pairs do contribute to
the long-range correlations. The way they do it reveals a
conspiracy between $\Phi_0$ and $\varphi$ \cite{Kosterlitz}. Due
to statistical independence of the fields $\Phi_0$ and $\varphi$,
the one-particle density matrix at large distances,
\begin{equation}
\rho(\mathbf{r})\, \propto\; \langle \; \exp[i \Phi(\mathbf{r}) -
i \Phi(0)]\; \rangle \; , \label{rho}
\end{equation}
 factorizes: $\rho(\mathbf{r})\,  \propto\,
\Gamma_{\Phi_0}(\mathbf{r})\,  \Gamma_{\varphi}(\mathbf{r})$,
where
\begin{equation}
\Gamma_{\Phi_0}(\mathbf{r})\;  =\;  \langle\;  \exp[i
\Phi_0(\mathbf{r}) - i \Phi_0(0)] \; \rangle \; , \label{}
\end{equation}
\begin{equation}
\Gamma_{\varphi}(\mathbf{r})\;  =\;  \langle\;  \exp[i
\varphi(\mathbf{r}) - i \varphi(0)] \; \rangle \; . \label{}
\end{equation} Remarkably,
the independent correlation functions $\Gamma_{\Phi_0}$ and
$\Gamma_{\varphi}$ make no physical sense separately, since they
both depend on the bare superfluid density $n_0$, but when the two
are combined in $\rho(\mathbf{r})$, the parameter $n_0$ drops out,
and the density matrix decays with the proper exponent $mT/2\pi
n_s$ \cite{Kosterlitz}.

A rational explanation of this ``secret agreement'' between
$\Phi_0$ and $\varphi$ is that being mathematically independent,
the two fields are deeply connected physically. The
above-mentioned structure of the correlation function even
suggests a qualitative form of the connection: The long-wave part
of the vortex-pair field $\Phi_0$ actually {\it belongs to
phonons, not vortices}.

In what follows, we trace the model (\ref{A}) back to its
dynamical Hamiltonian form and derive the canonical
parametrization of vortices and phonons from that starting point.
In doing so, we utilize the formalism recently developed by two of
us \cite{SvK_vortex_phonon}, from which it is directly seen that
the positions of vortices $\{\mathbf{r}_j\}$ and the field
$\varphi$ are canonical variables {\it only} in the limiting cases
of incompressible fluid ($\varphi\equiv0$) and vortex-free
environment ($\Phi_0 \equiv 0$) respectively. An explicit
transform from $\{\mathbf{r}_j\}$ and $\varphi(\mathbf{r})$ to the
canonical variables justifies that the deviation of the Jacobian
from unity is irrelevant in the context of the KT theory. Our
analysis allows us to reformulate KT theory in terms of canonical
variables: vortex pairs and genuine phonons, coupled to each other
in the most intuitive way: Vortex pairs interact with the
long-wave fluctuations of the velocity field precisely the same
way they interact with a homogeneous velocity field. This
interaction naturally accounts for the renormalization of the
long-wavelength-phonon stiffness and leads to the coarse-grained
effective action in the form of Eq.~(\ref{A_varphi}). In complete
agreement with physical understanding of vortex pairs as
essentially local objects, we demonstrate that the far-field of
$\Phi_0(\mathbf{r})$ belongs to phonons. Correspondingly, the
long-range decay of phase correlations is governed by the
statistics of long-wavelength phonons only, i.e. by the effective
action (\ref{A_varphi}).

Putting aside the issue of the Jacobian and {\it pior} to the
discussion of the dynamic model, a purely statistical insight into
the problem can be obtained by constructing an alternative to
$(\Phi_0, \varphi)$ set of variables. [For simplicity, below we
deal with only one vortex-anti-vortex pair; the generalization to
finite (but small) density of pairs is straightforward.] Consider
a vortex pair of separation ${\bf R}={\bf r}_1-{\bf r}_2$ located
at the point ${\bf r}_p=({\bf r}_1+{\bf r}_2)/2$, where ${\bf
r}_1$ and ${\bf r}_2$ are the positions of the vortices with
$l_1=1$ and $l_2=-1$ respectively. Introduce an auxiliary field
$\varphi_0({\bf r})$ such that it approaches $\Phi_0({\bf r})$
when $|{\bf r}-{\bf r}_p| \gg R$, and, in contrast to $\Phi_0({\bf
r})$, is regular at all distances. This definition fixes the
dipole moment of the new field:
\begin{equation}
 \int {\bf r}\, \Delta \varphi_0({\bf r}) \, d^2r \; =\; 2\pi ( {\bf R} \times \hat{z} ) \; , \label{}
\end{equation}
where $\hat{z}$ is a unit vector along $z$-axis. Make a
transformation
\begin{equation}
\varphi({\bf r})\; =  \; \tilde{\varphi}({\bf r}) \, -\,
\varphi_0({\bf r}) \; , \label{}
\end{equation}
which just shifts the field $\varphi$ by a regular $({\bf
r}_p,{\bf R})$-dependent field $\varphi_0$, and thus does not
change the configurational volume: ${\cal D}\varphi = {\cal
D}\tilde{\varphi}$. After this transformation, the long-range
behavior of the density matrix is completely described in terms of
the filed $\tilde{\varphi}$:
\begin{equation}
\rho(\mathbf{r})\, \propto\;  \langle \; \exp[i
\tilde{\varphi}(\mathbf{r}) - i \tilde{\varphi}(0)]\; \rangle \; .
\label{rho2}
\end{equation}
This simplification comes at a price: the vortex pair now couples to
$\tilde{\varphi}$. The structure of the interaction term between the
pair and the long-wave harmonics of the field $\tilde{\varphi}$
(such that $\lambda \gg R$, where $\lambda$ is the characteristic
wavelength) is most transparent. It reads
\begin{equation}
A_{\rm int}\; =\; {2 \pi n_0\over mT}\,  ( {\bf R}\times
 \hat{z})\cdot \nabla \tilde{\varphi} \Bigr|_{\mathbf{r}_p} \; ,
\label{}
\end{equation}
i.e. the vortex pair interacts with the long-wave part of
$\tilde{\varphi}$ exactly the same way it would interact with a
homogeneous velocity flow $(1/m)\,  \nabla \tilde{\varphi}
\Bigr|_{\mathbf{r}_p}$. One does not have to take this interaction
into account explicitly in the KT renormalization group treatment,
since its only relevant effect is to replace $n_0$ with $n_s$ for
phonons. [The effect of phonons on the statistics of vortex pairs is
negligibly small in the limit of $R\to \infty$, as is clear from a
direct estimate, see also below.] There is little doubt at this
point that $\tilde{\varphi}$ corresponds to genuine phonons and one
just needs to formally demonstrate this fact.

We start with Popov's hydrodynamic Lagrangian \cite{Popov},
\begin{gather}
L = \int \! \! \! d^2r \left[- n_0 \dot{\Phi}_0 -
\eta\dot{\varphi} - \eta \dot{\Phi}_0 \right] - H\; ,
\label{Lagrangian} \\
H= \int \! \! \! d^2r \Bigl[ \frac{n_0}{2m} \bigl| \nabla \varphi
\bigr|^2 + \frac{1}{2 \varkappa}\eta^2
\;\;\;\;\;\;\;\;\;\;\;\;\;\;\;\;\;\;\;\;\;\;\;\;\;\;\;\; \notag \\
\;\;\;\;\;\;\;\;\;\;\;\;\;\;\;\;\;\;\;\;\;\;\;\;\;\;\;\; +
\frac{n_0}{2m} \bigl| \nabla \Phi_0 \bigr|^2  + n_0 \mathbf{v}_0
\cdot \nabla \Phi_0 \Bigr]\; . \label{H_hydrodynamic}
\end{gather}
Here, the energy functional $H$ has been expanded to the leading
order with respect to small density fluctuations $\eta \ll n_0$,
$\Phi_0$ and $\varphi$ are defined by
Eqs.~(\ref{circulation})-(\ref{Laplacian_Phi0}), $\mathbf{v}_0$ is
the velocity of a global flow, and $\varkappa$ is the
compressibility. The typical vortex core size $\sim a_0 =
\sqrt{\varkappa/n_0 m}$ is much smaller than any other physical
length scale.

If the term $\int \! d^2r \, \eta \dot{\Phi}_0 \equiv T$ were
absent, $\eta$ and $\varphi$ would be the canonical conjugate
phonon variables, while
\begin{equation}
\int \! d^2r \, n_0 \dot{\Phi}_0\; =\; - 2\pi n_0 \sum_j l_j y_j
\dot{x}_j\; ,
 \label{}
\end{equation}
where $(x_j,y_j)\equiv\mathbf{r}_j$, would imply that $x_j$ and
$y_j$ are the canonical conjugate vortex variables. However, $T$ is
linear in the time derivatives of $x_j$ and $y_j$ and also contains
$\eta$ making the set of variables $\{\eta, \varphi \}$,
$\{\mathbf{r}_j\}$ not canonical, and meaning that $H$ in these
variables is not a Hamiltonian.

We are interested here only in the KT theory for the superfluid
phase in the vicinity of the transition, including the critical
point, where the concentration of vortex pairs of size $\sim R$ is
much smaller than $R^{-2}$ as $R\to \infty$. Correspondingly, at a
phonon wavelength $\lambda$ only pairs with $R\ll \lambda$
contribute to the renormalization of the sound velocity. This allows
us to use the small parameter
\begin{equation}
 R/\lambda \ll 1 \; \label{}
\end{equation}
for deriving canonical variables in the form of a regular
perturbative expansion starting from the zeroth approximation
$\{\eta, \varphi, \mathbf{r}_j\}$ \cite{SvK_vortex_phonon}.

It is straightforward to show that
\begin{equation}
 T = \sum_j 2 \pi l_j \;  [ \mathbf{\hat{z}} \times \nabla
Q(\mathbf{r}_j) ] \cdot \dot{\mathbf{r}}_j \; , \label{T1}
\end{equation}
where $Q(\mathbf{r})$ is defined by $\Delta Q (\mathbf{r}) =
\eta(\mathbf{r})$. We first switch to the Fourier representation
of $\{\eta, \varphi \}$:
\begin{gather}
\eta(\mathbf{r}) = \sum_{\mathbf{q}} \sqrt{\omega_{\mathbf{q}}
\varkappa/2V} \left[ \, e^{i \mathbf{q} \mathbf{r}} \,
c_{\mathbf{q}} + e^{-i \mathbf{q} \mathbf{r}}\, c^{*}_{\mathbf{q}}
\, \right] \; , \notag
\\ \varphi(\mathbf{r}) = - i \sum_{\mathbf{q}} \sqrt{1 / 2 V\omega_{\mathbf{q}} \varkappa }
\left[ \, e^{i \mathbf{q} \mathbf{r}} \, c_{\mathbf{q}} - e^{-i
\mathbf{q} \mathbf{r}}\, c^{*}_{\mathbf{q}} \, \right]\; ,
\label{phonon_fields}
\end{gather}
where $\omega_{\mathbf{q}}=(\sqrt{n_0/\varkappa \, m}) \, q$ and
$\{ c_\mathbf{q}, c^*_\mathbf{q} \}$ are resembling (and to the
zeroth approximation coincide with) the classical-field
counterparts of phonon creation and annihilation operators, and
$V$ is the system volume. Let the vortex ($l_1=1$) and the
anti-vortex ($l_2=-1$) in a pair have coordinates $\mathbf{r}_{1}
= \mathbf{r}_p + \mathbf{R}/2$ and $\mathbf{r}_{2} = \mathbf{r}_p
- \mathbf{R}/2$ respectively. Next, we expand $Q(\mathbf{r}_j)$ in
$T$ into series with respect to $q R \ll 1$. The resulting terms
are eliminated by iteratively correcting the variables
$\{\mathbf{r}_j \}$, $\{c_{\mathbf{q}},c^{*}_{\mathbf{q}} \}$ as
described in Ref.~\cite{SvK_vortex_phonon}, so that the Lagrangian
takes on the canonical form
\begin{equation}
L = \sum_{\mathbf{q}} i \, \dot{\tilde{c}}_{\mathbf{q}}
\tilde{c}^{*}_{\mathbf{q}} + 2 \pi n_0 \sum_j l_j \tilde{y}_j
\dot{\tilde{x}}_j  - H
\{\tilde{\mathbf{r}}_j,\tilde{c}_{\mathbf{q}},
\tilde{c}^{\dagger}_{\mathbf{q}}\}\; , \label{}
\end{equation}
where $\tilde{\mathbf{r}}_j=(\tilde{x}_j, \tilde{y}_j)$ and
$\tilde{c}_{\mathbf{q}}$,$\tilde{c}^*_{\mathbf{q}}$ are the
Hamiltonian variables. For our purposes we shall need only the
leading correction, which does not change the vortex variables,
\begin{equation}
\mathbf{r}_j\,  =\,  \tilde{\mathbf{r}}_j\; ,
\label{r_j_canonical}
\end{equation}
and for the phonon variables, yields
\begin{equation}
c_{\mathbf{q}}\;  =\; \tilde{c}_{\mathbf{q}}\,  +\,  2 \pi
\sqrt{\frac{n_0\, a_0\, q}{2V_2}} \; \frac{ [ q_x \tilde{R}_y -
q_y \tilde{R}_x ]}{q^2} \; e^{i \mathbf{q}\tilde{\mathbf{r}}_p }\;
, \label{c_q_canonical}
\end{equation}
where an equivalent set of vortex variables, $\tilde{{\bf
R}}=\tilde{{\bf r}}_1-\tilde{{\bf r}}_2$, $\tilde{{\bf
r}}_p=(\tilde{{\bf r}}_1+\tilde{{\bf r}}_2)/2$, is used. The
Jacobian of the transformation
(\ref{r_j_canonical})-(\ref{c_q_canonical}) equals unity. If
higher-order (in $\eta/n_0 \ll 1$ and $qR \ll 1$) terms are
included in Eqs.~(\ref{r_j_canonical})-(\ref{c_q_canonical}), the
deviation of the Jacobian from unity is of the order of
$(\eta/n_0)(qR)$. From now on we omit the tildes over the vortex
canonical variables in view of Eq.~(\ref{r_j_canonical}).

The canonical phonon fields, $\tilde{\eta}$, $\tilde{\varphi}$, are
defined analogously to (\ref{phonon_fields}), with
 $\{\tilde{c}_{\mathbf{q}},\tilde{c}^{*}_{\mathbf{q}} \}$ replacing
 $\{c_{\mathbf{q}},c^{*}_{\mathbf{q}} \}$. After substituting
(\ref{c_q_canonical}) for $c_{\mathbf{q}}$ in
(\ref{phonon_fields}) and taking the sum over $\mathbf{q}$ the
original variable $\varphi$ is expressed in terms of the canonical
variables as
\begin{equation}
\varphi(\mathbf{r}) = \tilde{\varphi}(\mathbf{r}) -
\frac{[(\mathbf{r}-\mathbf{r}_p)_y R_x -
(\mathbf{r}-\mathbf{r}_p)_x R_y]}{|\mathbf{r}-\mathbf{r}_p|^2}\; .
\label{varphi}
\end{equation}
Now we note that at large distances from the vortex pair
$|\mathbf{r}-\mathbf{r}_{p}| \gg R$, the field $\Phi_0(\mathbf{r})$
is given by \cite{Kosterlitz}
\begin{equation}
\Phi_0(\mathbf{r}) = \frac{[(\mathbf{r}-\mathbf{r}_p)_y R_x -
(\mathbf{r}-\mathbf{r}_p)_x R_y]}{|\mathbf{r}-\mathbf{r}_p|^2}\; .
\label{Phi_0_pair}
\end{equation}
Along with Eq.~(\ref{varphi}), this implies that sufficiently far
from $\mathbf{r_p}$, $\Phi_0(\mathbf{r})$ does not belong to the
vortex-anti-vortex pair at all, but is actually a part of the
phonon field $\tilde{\varphi}$.

After a standard algebra, the Hamiltonian assumes the form:
\begin{gather}
H=H_{\mathrm{v}} + H_{\mathrm{ph}} + H_{\mathrm{int 1}}+ H_{\mathrm{int 2}}\; , \label{H_pair} \\
H_{\mathrm{v}} = \frac{2 \pi n_0}{m} \ln (R/l_0) + 2 E_c,
\notag \\
H_{\mathrm{ph}} = \int \! \! \! d^2 r \Bigl[ \frac{n_0}{2m} \bigl|
\nabla \tilde{\varphi} \bigr|^2 + \frac{1}{2
\varkappa}\tilde{\eta}^2 \Bigr], \notag \\
H_{\mathrm{int1}} = 2\pi n_0 \; (\mathbf{R}
\times \hat{z}) \cdot \mathbf{v}_0, \notag \\
H_{\mathrm{int2}} = \frac{2\pi n_0}{m} \; (\mathbf{R} \times
\hat{z}) \cdot \nabla \tilde{\varphi} \Bigr|_{\mathbf{r}_p}.
\notag
\end{gather}
This form  is almost identical to the original effective action in
terms of $\mathbf{R}$ and $\varphi$ of Refs.~\cite{KT, Kosterlitz,
Jose, Nelson_Kosterlitz}. Besides the presence of $\eta$, which is
trivially integrated out in the partition function, the only
distinctive feature of the Hamiltonian (\ref{H_pair}) is the term
$H_{\mathrm{int2}}$, which couples the vortex dipole moment
$\mathbf{R}$ to the fluid velocity in the sound wave $\propto
\nabla \tilde{\varphi} \bigr|_{\mathbf{r}_p}$.

Consider the thermodynamics of the system (\ref{H_pair}) near the
critical point $T_c=\pi n_s/2 m$. The coupling term
$H_{\mathrm{int2}}$ does not change the statistics of vortices,
since its typical value is small, $H_{\mathrm{int2}}/T \sim q R
\lesssim 1$, whereas the contribution of $H_{\mathrm{v}}$ diverges
logarithmically. Therefore, the superfluid density, $n_s=(1/m
V)\;\partial^2 F/\partial v^{2}_{0
\alpha}\bigr|_{\mathbf{v}_0=0}$, $\alpha=x,y$, where $F=-T \ln Z$,
is given by the Kosterlitz renormalization group flow
\cite{Kosterlitz}. In contrast, $H_{\mathrm{int2}}$ is essential
for the phonon statistics. Since its structure is identical to the
vortex coupling to the uniform flow, averaging over $\mathbf{R}$
and $\mathbf{r}_p$ straightforwardly leads to the coarse-grained
effective action for the long-wavelength phonons governed by the
renormalized stiffness, Eq.~(\ref{A_varphi}).

To summarize, we have shown that the simplicity of the
parametrization (\ref{Phi})-(\ref{Laplacian_Phi0})---the statistical
independence of the fields $\Phi_0$ and $\varphi$---comes at a price
of substantial lack of its physical meaning, apart from
inconvenience of calculating off-diagonal correlators, where direct
contribution of vortex pairs has to be explicitly evaluated with the
only goal to replace bare superfluid density $n_0$ with its
renormalized value. An alternative parametrization in terms of
phonon variables (or their XY equivalents) renders the
Kosterlitz-Thouless scheme even more mathematically simple and
accurate, while making it physically transparent. The vortex-phonon
interaction that appears in the effective Hamiltonian does not lead
to any complications, because the structure of this interaction is
exactly the same as that of the interaction of vortex pairs with a
homogeneous external flow and the only effect of this interaction is
to ensure that {\it both} phonons and vortices are controlled by the
renormalized superfluid density.

This work was supported by the National Science Foundation under
Grants Nos. PHY-0426881, PHY-0456261, and by the Sloan Foundation.

\end{document}